# HearFit$^+$: Personalized Fitness Monitoring via Audio Signals on Smart Speakers

Yadong Xie, *Graduate Student Member, IEEE,* Fan Li, *Member, IEEE,* Yue Wu, and Yu Wang, *Fellow, IEEE*

**Abstract**—Fitness can help to strengthen muscles, increase resistance to diseases, and improve body shape. Nowadays, a great number of people choose to exercise at home/office rather than at the gym due to lack of time. However, it is difficult for them to get good fitness effects without professional guidance. Motivated by this, we propose the first personalized fitness monitoring system, HearFit$^+$, using smart speakers at home/office. We explore the feasibility of using acoustic sensing to monitor fitness. We design a fitness detection method based on Doppler shift and adopt the short time energy to segment fitness actions. Based on deep learning, HearFit$^+$ can perform fitness classification and user identification at the same time. Combined with incremental learning, users can easily add new actions. We design 4 evaluation metrics (i.e., duration, intensity, continuity, and smoothness) to help users to improve fitness effects. Through extensive experiments including over 9,000 actions of 10 types of fitness from 12 volunteers, HearFit$^+$ can achieve an average accuracy of 96.13% on fitness classification and 91% accuracy for user identification. All volunteers confirm that HearFit$^+$ can help improve the fitness effect in various environments.

**Index Terms**—Fitness monitoring, user identification, acoustic sensing, smart speaker.

✦

## 1 INTRODUCTION

NOWADAYS, more and more people pay attention to their health status and body shape, causing a rapidly growing popularity of fitness. Effective fitness can bring numerous benefits, such as increasing muscle strength, improving body shape, and decreasing risks of cardiovascular diseases [1], [2]. However, due to the fast pace of life, it is inconvenient for many people to go to dedicated gyms. Besides, it usually costs much to become a gym membership and consult personal trainers. Therefore, plenty of people begin to do fitness at home/office, which is more conducive to save time and money. But due to lack of effective supervision and professional effect evaluation, they usually get unsatisfactory fitness effects.

The usual way to get fitness instruction is to hire a personal trainer, which is not suitable at home/office. There are several fitness APPs on the market [3]–[5] that can provide a few fitness guidance. However, these APPs can neither monitor fitness processes in real-time, nor provide fine-grained fitness statistics. In order to facilitate people to exercise at home/office, researchers propose many solutions. Some fitness monitors are realized by analyzing videos [6]–[8], but they depend on good lighting conditions and may involve privacy issues. There is a new trend of using wearable devices for fitness monitoring. For example, RecoFit [9] uses inertial sensors worn on forearms to track strength training. However, these devices need to be worn all the time during fitness, which adds an extra burden to users and makes them uncomfortable. Thus, some researchers use RF signal for non-invasive fitness monitoring. TTBA [10] monitors free-weight exercise by attaching RFID tags on dumbbells and using Received Signal Strength Indication (RSSI) to recognize the motion types, but it can only monitor exercises that require dumbbells. FitAssist [11] can recognize exercise and assess quality based on Wi-Fi devices. However, the Wi-Fi-based approaches usually need a pair of antennas, and users need to exercise between them. Besides, Wi-Fi signal is easy to be interfered by environmental factors like the movements of other people. Therefore, it is an urgent need to develop a non-invasive, easy-to-deploy, and anti-interference fitness monitoring system to provide fine-grained fitness statistics and help to improve fitness effect.

Toward this end, we take one step forward to study the feasibility of using acoustic sensing, which achieves great success in human activity recognition [12]–[15], for fitness monitoring. In addition, smart speakers like Amazon Echo and Google Home are becoming popular. As of 2019, there are 157 million smart speakers in U.S. homes and an average of 2.6 smart speakers in each household with smart speakers [16], which bring great potential in fitness monitoring without attaching any devices to users. In this paper, we design and implement HearFit$^+$, the first personalized fitness monitoring system based on commercial smart speakers for home/office environments. HearFit$^+$ aims to extract features that can perform user identification and provide users with fine-grained fitness statistics to improve fitness effect in a more comprehensive method.

The key idea of HearFit$^+$ is that the smart speaker emits inaudible ultrasonic signal, the signal is reflected by the user and then received by the microphone array, which is then processed to extract the user's identity and fitness statistics. However, there are multiple challenges to realize HearFit$^+$. Firstly, people perform non-fitness activities most of the day,

• Y. Xie, F. Li and Y. Wu are with the school of Computer Science and Technology, Beijing Institute of Technology, Beijing, 100089, P.R.China. E-mail: {ydxie, fli, ywu}@bit.edu.cn
• Y. Wang is with the Department of Computer and Information Sciences, Temple University, Philadelphia, Pennsylvania 19122, USA. E-mail: wangyu@temple.edu
• F. Li is the corresponding author.

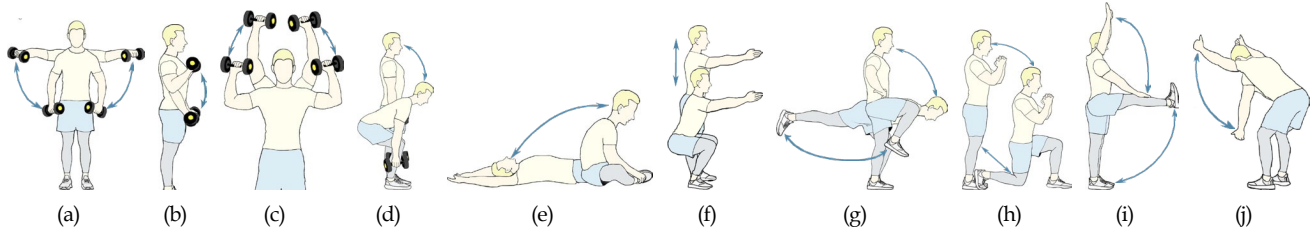

Fig. 1. Illustration of 10 types of fitness actions: (a) Lateral raise; (b) Dumbbell curl; (c) Shoulder press; (d) Dumbbell deadlift; (e) Sit-up; (f) Squat; (g) One leg deadlift; (h) Lunge; (i) Front kick; (j) Y stretch.

so HearFit+ needs to distinguish fitness from daily activities automatically. Secondly, at home/office, the reflected signal of the exercising user is easily interfered by the activities of surrounding people (e.g., walking or typing), which requires HearFit+ to be anti-interference. Finally, HearFit+ not only determines the identity of the user, but also classifies and evaluates the fitness actions. In addition, HearFit+ needs to have high scalability, which allows users to add new actions.

To address these challenges, we first analyze the fitness activities and find that they are typically more repetitive than non-fitness activities. Thus, we use the autocorrelation of acoustic signal to determine the activity mode of the user. In order to minimize the interference of surrounding people, we use the microphone array to locate the user. But the close distance between each microphone (i.e., about $5cm$) makes it impracticable to calculate an accurate distance between the user and the smart speaker through triangulation. So we calculate the accurate direction of the user by Generalized Cross-Correlation with phase transform (GCC-PHAT) [17] and use beamforming to amplify the reflected signal of the user. Finally, we design a method to segment each fitness action based on Short Time Energy (STE) [18], and design a Long Short-Term Memory (LSTM) network to classify the type of each action. User identification is challenging because the same fitness actions of different users are very similar. Fortunately, we find that there are unique micro-action patterns of different users that can be captured from the LSTM network. Besides the actions that are trained in advance, users can add new actions. Combined with incremental learning [19], HearFit+ only needs to collect a small training set of new actions to train the last few layers of the neural network. To evaluate the fitness effect, we study the Frequency, Intensity, Time and Type (FITT) principle [20] commonly used in fitness evaluation, and define 4 metrics to evaluate the fitness process from local view and global view.

The prototype of HearFit+ is built using a Raspberry Pi 4B, a circular 6-microphone array, and an omni-directional speaker, which is the same as most of smart speakers in the market. To evaluate the performance of HearFit+, we recruit 12 volunteers (9 males and 3 females) and record their fitness process in 4 different rooms. The experiments involve 10 typical types of fitness, as shown in Fig. 1. Finally, we collect over 9, 000 actions of all types of fitness. All procedures are approved by the Institutional Review Board (IRB) at Beijing Institute of Technology. Results demonstrate that HearFit+ can accurately identify both fitness types and users in various indoor environments, and provide fine-grained fitness statistics.

Our contributions are summarized as follows:

- We propose a fitness monitoring system, HearFit+, using smart speakers at home/office. It can provide users with fine-grained fitness statistics and help users to improve fitness effects in a comprehensive method. To the best of our knowledge, we are the first to use smart speakers for personalized fitness monitoring.
- We present several methods to extract features from reflected signal. We design a novel method to segment actions. And compared with the existing research using the LSTM network, we innovatively modify the structure of the LSTM network to enable one LSTM network to achieve two functions (i.e., fitness classification and user identification). We define 4 metrics from local and global views to help to improve fitness effects.
- Using the characteristics of the microphone array, HearFit+ has good anti-interference ability. Combined with incremental learning, users can easily add new actions to the system.
- We evaluate HearFit+ using a hardware prototype and conduct extensive experiments with 12 volunteers in different environments. The results show that HearFit+ can achieve an average accuracy of 96.13% on fitness classification and 91% accuracy on user identification, and provide accurate fitness effect evaluation.

The remainder of the paper is organized as follows. We review related work in Section 2. Section 3 presents the detailed design of HearFit+. Implementation and extensive experimental results are provided in Section 4. We discuss the limitations and future work in Section 5. Finally, we draw our conclusion in Section 6.

## 2 RELATED WORK

Recent works show that fitness monitoring can be achieved by various technologies, including camera-based, wearable device-based, and wireless device-based methods. Furthermore, we also review identification based on acoustic sensing.

**Camera-based methods.** Some research leverages cameras for fitness monitoring and activity recognition. Fitness Mate [6] uses Microsoft Kinect to monitor and guide physical exercise. MotionMA [21] extracts models of movement demonstrated by users and provides real-time feedback on how to improve their performance by using Kinect. A Hidden Semi-Markov Model-based approach [22] monitors and evaluates body motions during rehabilitation training by extracting features from skeleton joint trajectories acquired by the RGB-D camera. These methods, however, highly depend on good lighting conditions. And people usually refuse to deploy cameras at home/office due to privacy considerations [23].

**Wearable device-based methods.** The development of wearable devices provides a new way for fitness monitoring. iCoach [24] is a smart fitness glove with inertial measurement units, which can recognize exercise and assess workout quality. FitCoach [25] leverages smartphones or smartwatches mounted on upper arms to achieve fine-grained fitness interpretation and smart exercise guidance. GymSoles [26] is an insole prototype that provides feedback on the centre of pressure at the feet to assist users with maintaining the correct posture while performing squats and dead-lifts. However, these methods require users to wear additional devices and can only recognize the fitness of the body part where the device is attached.

**Wireless device-based methods.** There are also some works using wireless devices to recognize fitness. FEMO [27] monitors fitness by attaching passive RFID tags on the dumbbells and adopts Doppler shift for action recognition and assessment. But, it can only monitor exercises that require dumbbells and users need to prepare dedicated RFID readers. Motion-Fi [28] can count repetitive motions and enable multi-users to perform motions leveraging wireless backscattering. SEARE [29] is a fitness monitor which provides users with health management during exercise using CSI-waveform shape as features. But the Wi-Fi-based approaches usually need a pair of antennas, and users need to exercise between the antennas. Besides, Wi-Fi signal is easy to be interfered by environmental factors like movements of other people [30]. At present, only a few works [31], [32] use ultrasound to achieve fitness monitoring. But they only classify 3 types of fitness. They extract short-time Fourier transform (STFT) from the reflected ultrasound of fitness actions and test different classification schemes to evaluate the classification results. However, they do not design noise reduction, segmentation, and effect evaluation algorithms, so it is very difficult for them to deploy in actual use environments.

**Identification using acoustic sensing.** Some research uses active acoustic sensing for user identification. Lip-Pass [15] extracts unique behavioral characteristics of users' speaking lips leveraging built-in audio devices on smartphones for user identification. TouchPrint [33] performs active acoustic sensing to extract the fine-grained acoustic multipath effect features corresponding with the hand geometry and postural information when the finger is static on the touch screen for user identification. VocalLock [34] is a user identification system, which senses the whole vocal tract during speaking to identify different individuals in a passphrase-independent manner on smartphones leveraging acoustic signals. Different from these research, we explore the feasibility of using acoustic sensing to identify the individual who performs various fitness activities.

We are the first to achieve non-invasive fitness monitoring and user identification at the same time using acoustic sensing [12]–[15]. Unlike the above works, HearFit+ automatically monitors fitness and evaluates fitness effects of users without wearing extra devices. Through taking advantages of the omni-directional speaker and microphone array, HearFit+ has good anti-interference ability. Compared with our early conference version of this work [35], we mainly enhance feature extraction and fitness classification. Specifically, we further reduce the high-dimensional features to two dimensions. The results intuitively show that the Doppler shifts of different actions and users are unique. In addition, we modify the structure of the network so that it can identify users. In the experimental part, we conduct 4 new experiments to verify the performance of our newly proposed algorithm.

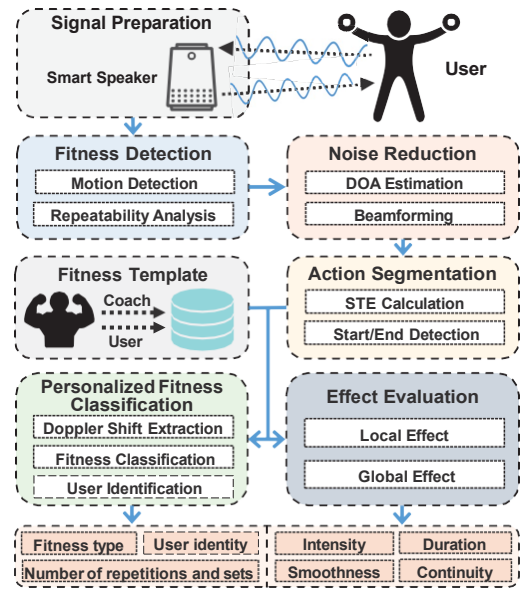

Fig. 2. System architecture of HearFit+.

## 3 SYSTEM DESIGN

This section details the design of HearFit+ and highlights the core techniques behind the segmentation, fitness classification, user identification and fitness evaluation.

### 3.1 System Overview

Fig. 2 shows the architecture of HearFit+. It mainly includes 6 parts, which are *Signal Preparation*, *Fitness Detection*, *Noise Reduction*, *Action Segmentation*, *Personalized Fitness Classification*, and *Effect Evaluation*. In order to classify fitness actions and evaluate effects, we recruit professional fitness coaches in advance to collect standard fitness data as *Fitness Template*. After observing the coaches, we divide a fitness process into repetitions and sets. A repetition is one complete cycle of a fitness action, while a set usually contains a certain number of repetitions.

In *Signal Preparation*, we design an acoustic signal considering various factors. The speaker emits the signal, which is then reflected by surrounding objects and received by the microphone array. Note that here we get 6 signals from 6 microphones. Then, HearFit+ performs *Fitness Detection* to detect whether there are fitness activities around the smart speaker. Specifically, we adopt a band-pass filter on the reflected signal and search for Doppler shifts that may be caused by motions. If the Doppler shifts are repetitive, the corresponding motions are considered to be fitness actions. Once the fitness is detected, HearFit+ performs *Noise Reduction*. The Doppler shifts from all the 6 microphones are processed to get the Direction of Arrival (DoA) of signals. To increase the signal strength of the fitness and suppress noise, we use beamforming to synthesize 6 signals. In *Action Segmentation*, we calculate the STE to capture the energy pattern of the signal and analyze the slope of the STE to detect the start and end time of each repetition. By analyzing the time interval between each repetition, we can calculate the number of repetitions and sets.

HearFit+ determines the type of each repetition and the identity of the user in *Personalized Fitness Classification*. In detail, we perform Fast Fourier Transform (FFT) on the signal to extract effective features of Doppler shifts. The features are sent to a bi-functional LSTM network, which is trained by fitness data of coaches and users, to determine the fitness type and the user identity. After fitness, users usually care about fitness effects to ensure that muscles are strengthened and injury risks are low. So *Effect Evaluation* aims to provide feedback to users. The feedback includes two aspects: local effect and global effect. Local effect denotes the quality of each repetition by evaluating intensity and duration. The global effect denotes the continuity and smoothness of each set. Finally, HearFit+ provides users with the statistics to help them improve fitness effects. Through incremental learning, users can add new actions to *Fitness Template* with a cost of a small amount of training.

## 3.2 Signal Preparation

To monitor a user's fitness process, we turn the smart speaker into an active sonar and observe Doppler shifts to detect the user's motion. Doppler shift is the change in frequency of a signal in relation to an observer who is moving relative to the signal source [36]. Formally, the Doppler shift $\Delta f$ is determined by $\Delta f = (2v/c) \cdot f$, where $f$ is the emitted frequency, $c$ is the speed of sound and $v$ is the speed of relative movement. In HearFit+, the smart speaker is not only the signal source but also the observer. The user who reflects the signal can be seen as a virtual signal source. To accurately detect the user's motion, the emitting signal needs to be carefully designed.

In indoor environments, the larger the bandwidth of the signal, the less the interference of surrounding objects, such as roof, wall, and furniture [37]. Thus, we design a wideband emitting signal by considering several factors. First, the signal should be inaudible to people. According to [38], the frequency above 18*kHz* is already inaudible to kids who are sensitive to sounds. Second, a higher frequency of a sound results in a more obvious Doppler shift. However, a related research [39] shows that the signal attenuates with the increase of sound frequency due to the influence of hardware performance, namely the sharp anti-aliasing filter. We use multiple speakers and microphones to perform experiments at different frequencies, and the results show that the attenuation is acceptable when the frequency is lower than 21*kHz*, which is supported by most of the speakers [40]. Third, frequency selective fading may occur with a single frequency due to the multipath effect of the signal, which greatly decreases the system performance. It can be reduced by emitting signal with multiple frequencies. Therefore, comprehensively considering the factors, we design a signal with 5 frequency components (i.e., 19*kHz*, 19.5*kHz*, 20*kHz*, 20.5*kHz*, and 21*kHz*). The average wavelength of the signal is about 1.7*cm*. As most of the obstacles (e.g., walls, floor, and users) present in a room are usually larger than this wavelength, we can consider that the reflected signal only includes specular reflection [41], which has strong energy. The signal is defined as:

$$s(t) = \sum_{n=1}^{5} A_n \sin(2\pi f_n t + \varphi_n), \quad (1)$$

where $f_n$ denotes the $n$-th frequency component. $A_n$ and $\varphi_n$ are the amplitude and initial phase, respectively. We select the 5 frequency components for two reasons. First, when users exercise, the speed of their movements usually does not exceed 1.5*m/s*, which can cause a Doppler shift of 185.2*Hz* under a sound of 21*kHz*. Since 185.2*Hz* is smaller than the half of intervals between frequency components, their Doppler shifts do not overlap in the frequency domain. Second, we find that more than 5 frequency components would make the signal audible even at a low volume due to sub-harmonics.

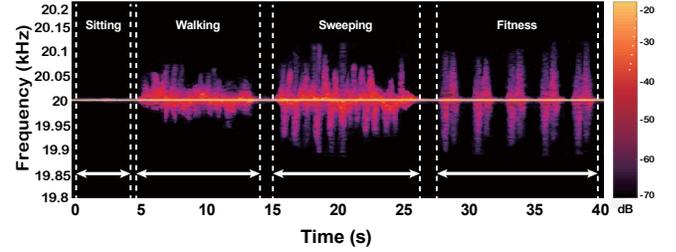

Fig. 3. Doppler profile of several activities.

At the receiving end, the 6-microphone array collects reflected signals at a sampling rate of 44.1*kHz*. Through the analysis of the collected data, we find that each type of fitness action has a unique pattern on its Doppler shift, which suggests that Doppler shift can be used in fitness monitoring.

## 3.3 Fitness Detection

HearFit+ first detects motions around the smart speaker. After that, it needs to distinguish between fitness actions and non-fitness activities (e.g., sitting, walking, and sweeping) contained in those motions. We first conduct several preliminary experiments on recordings of several human activities. For example, a user first sits, then walks to a location to sweep the floor, and finally performs 5 fitness actions. Fig. 3 shows the corresponding Doppler profile of 20*kHz* signal collected by one microphone. We can observe that fitness actions are more repetitive than non-fitness activities. This is because in order to improve fitness effects, users usually divide a fitness process into multiple sets, and each set contains multiple identical fitness actions with similar intervals between every two actions. Based on this observation, we design a two-step detection method to determine the start time of the fitness.

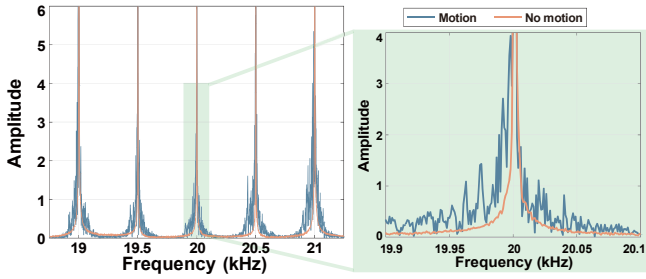

Fig. 4. Doppler shifts of motion and no motion.

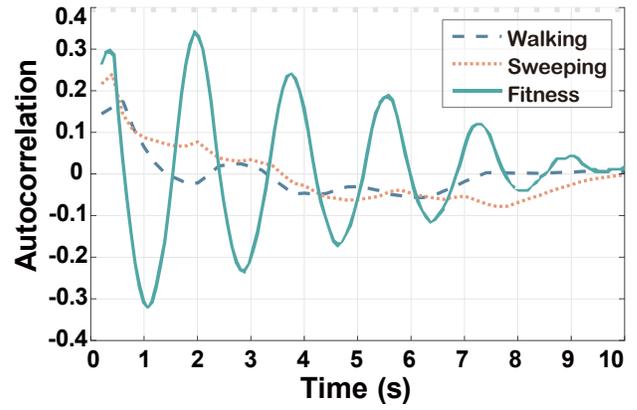

Fig. 5. Autocorrelation of 3 types activities.

**Motion detection.** First, we check whether there are motions around the smart speaker through analyzing the amplitudes of Doppler shifts. By observing Fig. 3, we can find that when the user is sitting, there is the reflected signal by static objects (e.g., roof and wall) without Doppler shifts. When the user has motions, the reflected signal contains Doppler shifts. So, to detect Doppler shifts caused by motions, HearFit+ emits one-second signal every $10s$, then FFT is performed on the signal received by each microphone. Fig. 4 shows the FFT results under conditions with motions and without motions. We can see that when the user has motions, there are several peaks with larger amplitude around the center of each frequency component. On the contrary, when there is no motion, the amplitude of each frequency component sharply raises and drops to a negligible value. Based on such observation, we extract two closest peaks on each side of each frequency component. Finally, we can obtain 120 peaks (i.e., 2 peaks × 2 sides × 5 frequency components × 6 microphones) for each one-second signal and calculate the average value of these peaks. If the average value exceeds a threshold, we consider that there exist motions. After extensive experiments, we set this threshold to 1.

**Repeatability analysis.** Once HearFit+ detects motions, it starts to emit consecutive signal. We add a window with a length of $13s$ that slides $1s$ each time on the reflected signal, and adopt a band-pass filter on each window to obtain the target frequency band ranging from $18.5kHz$ to $21.5kHz$. We then perform a spectral subtraction by subtracting the magnitude of the 5 frequency components from the filtered signal. Then we add the phase of the filtered signal to the subtraction result. All negative values are set to the noise floor before adding the phase, while only Doppler shifts are remained in the processed signal. HearFit+ further determines whether the motions are fitness activities by calculating autocorrelation [42] of the Doppler shifts. Autocorrelation is the correlation of a signal and its delayed copy. It can indicate whether a signal is periodic. If a signal has a strong periodic component, there is a peak in autocorrelation. On the contrary, the autocorrelation of irregular motions (with no periodicity) shows no peaks at all. We denote a signal as $X_t$ and its copy that delay for a period of $k$ as $X_{t+k}$, then its autocorrelation is:

$$R(k) = \frac{E[(X_t - \mu)(X_{t+k} - \mu)]}{\sigma^2}, \quad (2)$$

where $\mu$ and $\sigma$ are the mean and standard deviation of $X_t$,

respectively. Fig. 5 shows the autocorrelation of the 3 types of activities. We can see that fitness actions have obvious repetitive patterns, while non-fitness activities hardly have long-term repetitive patterns. Then, we count the peaks that have autocorrelation values larger than a threshold (empirically set as 0.1). If each of two consecutive windows both has more than 4 peaks, we consider there exist fitness actions.

HearFit+ can only adopt repeatability analysis without motion detection to detect fitness. But HearFit+ needs to emit consecutive signal and analyze data all the time. However, by observing usage scenarios, we find that there are usually no human activities around the smart speaker most of the time. Thus, we design motion detection that HearFit+ just emits and collects 1-second signal every $10s$ to make our system more energy-efficient. Thanks to the voice interaction function of smart speakers, we also allow users to use voice commands to start fitness monitoring to deal with some exceptional cases, such as HearFit+ does not detect motions correctly.

### 3.4 Noise Reduction

At home/office, the reflected signal from the fitness user is easily interfered by the activities of surrounding people (e.g., walking or typing on the computer). In order to minimize these interferences after HearFit+ detects fitness, we take advantage of the microphone array to direction-ally enhance the reflected signal of the fitness user. For convenience, we establish a suitable coordinate system, as shown in Fig. 6. For an array with $M$ microphones, the azimuth angle of the $m$-th microphones is given by $\phi_m = (-(M-1)/2 + m - 1) \cdot 360/M$. The azimuth angle is defined as the angle from the x-axis toward the y-axis in the xy-plane. The elevation angle is defined as the angle from the xy-plane toward the z-axis.

**DOA estimation.** First, we need to get the user's direction relative to the smart speaker. Considering the application scenarios, the signal emitted from the smart speaker is reflected by multiple objects (i.e., roof, wall, and user), but only the active user can make the reflected signal with Doppler shifts. After repeatability analysis, the reflected signals of the static objects are subtracted, remaining only the reflected signal of the user. At present, the commonly used array-signal localization algorithms include multiple signal classification (MUSIC) and GCC-PHAT. MUSIC is usually used to process stationary narrowband signals, but in our application scenarios, we collect non-stationary wideband audio signals. In addition, MUSIC algorithm needs to search the location of the sound source in the target space, which has high computational complexity. Compared with MUSIC, GCC-PHAT [17], [43] has less computation, which is conducive to real-time processing. It can be applied to non-stationary wideband audio signals. Therefore, we adopt GCC-PHAT algorithm to determine the user's location.

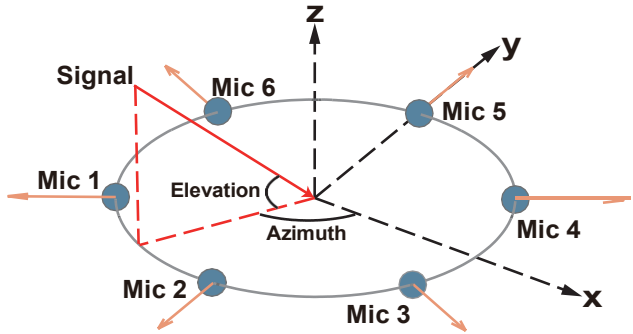

Fig. 6. Coordinate system of the microphone array.

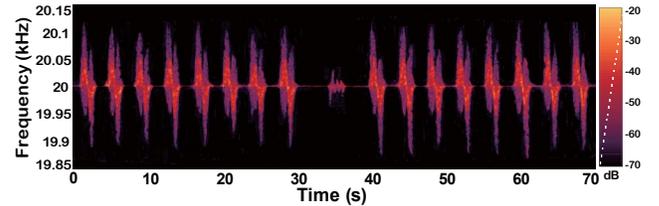

(a) Doppler profile of 16 lunges.

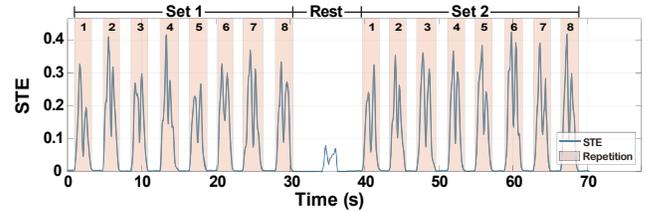

(b) Segmentation result of 16 lunges.

Fig. 7. The process of action segmentation.

GCC-PHAT assumes that signal source is located in the array far-field, so the DOA is the same for all microphones. We use the signals of the 6 microphones processed in repeatability analysis and estimate the correlations between each signal pair by GCC-PHAT and find the largest peak in each correlation. The peak identifies the delay between the two signals that arrive at microphones in a pair. Finally, a least-squares estimation is used to derive the azimuth and the elevation angles of the user. Since most users do not change their locations during fitness, the DOA estimation only needs to run once at the beginning of fitness.

**Beamforming.** Now we have an initial estimation of the azimuth and the elevation angles of the user. Then we use a time-delay beamforming algorithm to suppress interference from other people and increase the SNR of the reflections from the user. The time-delay beamforming compensates a reflected signal coming from a specific direction for the arrival time differences across the microphones. It includes two steps: time alignment and synthesis. Suppose that the azimuth and elevation angles of the user are $\alpha$ and $\beta$ respectively, the delay $\Delta_m$ of the $m$-th microphones can be calculated as $\Delta_m = r \cos(\alpha - \phi_m) \cos \beta / c$, where $r$ is the radius of the microphone array. The individual signals are then synthesized by $S(t) = \sum_{m=1}^{6} s_m(t - \Delta_m)$, where $s_m(t)$ is the sample of the $m$-th microphones at time $t$. Finally, we get one copy of synthesized Doppler shifts $S(t)$ from the signals of 6 microphones.

Most smart speakers such as the Amazon Echo have circular LED lights used for various functions, such as indicating the direction of the voice command. Such LED lights can also be used for indicating the direction of fitness. In order to improve the usability of HearFit+, we also allow users to use voice commands to start fitness monitoring. Under this condition, we determine the direction of the user by detecting the direction of the user's voice commands. In order to detect voice commands, the user needs to stand in the position of fitness and say a command. To get the DOA of voice commands, we use a low-pass filter to focus on human voice and then use GCC-PHAT to obtain the direction.

### 3.5 Action Segmentation

In order to accurately classify fitness actions, identify users and provide fitness effect evaluation, we segment each repetition. In other words, we need to detect the start time and end time of each repetition. We find that when the user moves, the energy of the Doppler shifts also changes, so we calculate the STE [18] of the signal and segment each repetition based on the STE.

**STE calculation.** STE is widely used in speech recognition to separate voiced and unvoiced speech segments. We first normalize the signal using the Min-Max Normalization. Then, we apply a sliding window with a length of 0.5$s$ that slides 0.1$s$ each time on the signal. We define the STE of time $t$ as:

$$E_t = \sum_{\delta = t-l+1}^{t} [s(\delta) w(t - \delta)]^2, \qquad (3)$$

where $s(\delta)$ is the amplitude of the signal at time $\delta$, $w$ is the Hanning window and $l$ is the length of the window. Fig. 7(a) shows the Doppler profile of two sets of lunges and Fig. 7(b) shows the corresponding STE. We further find a unique pattern of the STE: 1) at the beginning of a repetition, STE increases rapidly; 2) STE drops slightly when the repetition stops in the middle; 3) STE increases again when the user returns to the initial pose; 4) STE decreases sharply when the user finishes a repetition. Based on this pattern, we design a method to detect the start time and end time of each repetition.

**Start/End detection.** Detecting the start time and end time of each repetition not only helps HearFit+ to classify fitness actions and identify users, but also provides the basis for the subsequent effect evaluation. The common segmentation methods are peak detection [44] and KL divergence algorithm [27], but they are not applicable in our system. Peak detection usually segments repetitions by detecting peaks or troughs in sensing data. However, it is only applicable when there is no rest interval between consecutive repetitions. KL divergence algorithm uses the discrete probability distribution of each window to identify repetition and rest interval. But the window size has a great influence on system performance, as a small window may cause a repetition to be divided into several parts, while a large window may bring a large amount of computation. Since the durations of fitness actions are different, it is difficult to choose a suitable window size.

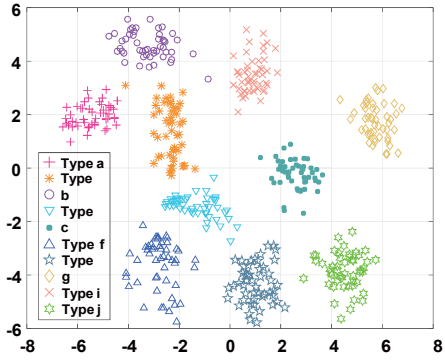

Fig. 8. t-SNE visualization of 10 typical fitness actions.

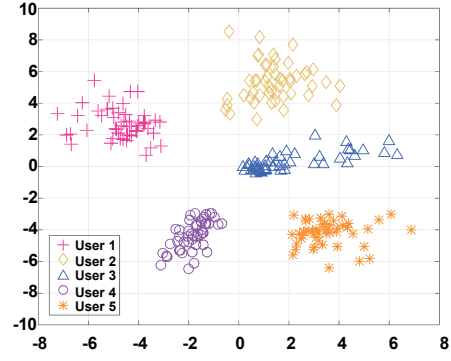

Fig. 9. t-SNE visualization of 5 different users.

Different from them, we segment each repetition in a simple but effective way based on the slope of STE. We set thresholds $sth$ and $eth$ for the slope of the start and end position, respectively. Since each repetition contains start and end time, we alternately search for the start and end time in the STE. We choose the time $t$ with STE $E_t$ less than 0.03, and calculate its slope by $\Theta_t = (E_{t+1} - E_t)/\Delta t$. If $\Theta_t$ is greater than $sth$, and $E_{t+2} > E_{t+1}$, we consider time $t$ as a start time. Then, we search for the corresponding end time after time $t$. If $\Theta_{t+\psi}$ is less than $eth$, and $E_{t+\psi-1} > E_{t+\psi}$, we consider time $t + \psi$ as corresponding end time. We then continue to search for the start and end time of the next repetition until the user finishes fitness. Fig. 7(b) shows the result of the segmentation, we can see that all start time and end time are correctly identified. Note that the result of the segmentation may include tiny movements during the rest between two sets, we filter out the tiny movements by analyzing the maximum energy of each repetition. Finally, if the time interval between two repetitions is more than $7s$, we consider that they belong to different sets. If there is no repetitive pattern within $90s$ after a repetition, we consider the fitness is finished.

### 3.6 Personalized Fitness Classification

After segmentation, HearFit+ aims to classify the fitness type of each repetition and identify the user. Considering that repetitions in a set are usually the same, HearFit+ only analyzes the first 3 repetitions in each set to reduce the amount of calculation. In the early stage, we recruit professional fitness coaches to collect standard fitness data as the fitness template. In order to achieve user identification, HearFit+ also needs to collect some personal fitness data of users in advance. These data are used to train a bi-functional LSTM network to classify the repetition and identify the user. Before training the network, we first extract effective features from these data.

#### 3.6.1 Doppler Shift Extraction

To accurately classify repetitions and identify the user, it is necessary to extract reliable features from each repetition's reflected signal. The frequency-domain information stored in the signal is widely used in activity recognition [45], [46]. In addition, each repetition is a consecutive action that lasts for a period, if we directly perform FFT on the signal, the time domain information is lost. So we design a time-frequency feature extraction method.

For the signal of each repetition, we further divide it into 8 blocks on average. To obtain frequency feature (i.e., Doppler shifts), each block is processed by a 4096-point FFT, which means dividing the sampling rate into 4096 points evenly. Since the frequency we use is mainly between $19kHz$ and $21kHz$, the information in other frequencies can be ignored. In other words, we just focus on the 1765th to 1950th points. Through experiments, we find that the amplitude of FFT results is better for training network than the phase. Thus, the amplitude of each block is used to form a 186-dimension feature vector, and each repetition gets 8 feature vectors.

We conduct a feasibility study to validate the effectiveness of the feature extraction method through t-distributed Stochastic Neighbor Embedding (t-SNE) [47], which is especially suitable for visualizing high-dimensional datasets. 5 users are asked to collect data of 10 typical fitness actions. We combine the 8 feature vectors of each repetition into a vector in order, and then use the t-SNE to reduce the dimension of the vector to 2. Specifically, we first label each repetition with the type of fitness and show the t-SNE visualization of the 10 typical fitness actions in Fig. 8. It can be seen from the figure that even if the fitness data come from different users, all 10 types of fitness actions are well separated. Then, we select a kind of fitness action randomly and label each repetition with the identity of the user. Fig. 9 shows the t-SNE visualization of the 5 different users. We can see from the figure that when different users perform the same fitness actions, their Doppler shifts are different. After analysis, we find that the micro-actions of different users are not exactly the same. Therefore, we need to design an accurate classifier to identify the difference

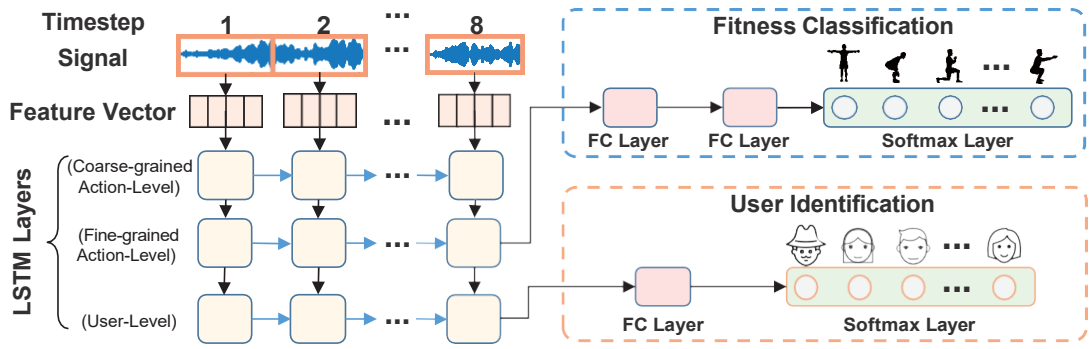

Fig. 10. Structure of the bi-functional LSTM network.

between different fitness actions and the difference between the micro-actions of different users.

### 3.6.2 Fitness Classification and User Identification

After getting feature vectors, we use a deep learning method to classify the fitness type of each repetition and identify the user. Traditional classifiers (e.g., k-NN, RF and SVM) usually treat each block as an independent unit and ignore the temporal context. However, each fitness action is not a transient action, but a consecutive action lasting for a time period. Therefore, ignoring the temporal context during modeling may limit the performance of our system in fitness classification and user identification. So, we design a bi-functional LSTM network [48] to exploit the temporal dependencies within blocks. The architecture of LSTM network is recurrent, which means that it considers not only the current block but also previous blocks. We set timesteps to 8 and each timestep takes a feature vector of a block as input. Finally, it classifies each repetition and associates the detected repetition with the corresponding user. Compared with existing research using LSTM network, our main innovation is to modify the structure of the LSTM network to enable one LSTM network to achieve two functions (i.e., fitness classification and user identification). The existing research usually uses two modules to achieve these two functions, which means more calculations and more complex algorithms need to be designed.

Fig. 10 shows the structure of the designed network, which has 3 LSTM layers, a fitness classification module, and a user identification module. The feature vector of each repetition is used as the input of the first LSTM layer, and the coarse-grained action-level features can be extracted as output by the compressed representations in the first layer. Then, the output of the first LSTM layer is fed to the second LSTM layer. The second layer further extracts the fine-grained action-level features. Finally, the third LSTM layer takes the output of the second layer as input, and extracts the user-level features (e.g., micro-action features), which represent the unique patterns of a user and can be used for user identification. Furthermore, to classify the repetition and facilitate users to add new actions, we add a fitness classification module with the feature representations from the second LSTM layer as the input. This module contains 2 fully connected (FC) layers and a softmax which can assign the repetition to the corresponding type of fitness action. After the third LSTM layer, we attach a user identification module with an FC layer and a softmax layer. This module can further determine the identity of the user.

LSTM layers are the most essential layer, each of which can transform the input to a set of compressed representations that can characterize unique fitness actions and users through an unsupervised manner. At $t$-th timestep, the LSTM layer can map the input $p_t$ into a compressed vector $h_t$ as follow:

$$h_t = \sigma \left( W_o \left[ h_{t-1}, p_t \right] + b_o \right) \cdot tanh \left( C_t \right), \quad (4)$$

where $\sigma(\cdot)$ is the sigmoid function, $W_o$ and $b_o$ denote a weight matrix and a bias vector of output gate, respectively. $C_t$ denotes the status at $t$-th timestep. We add 2 FC layers after the last timestep of the second LSTM layer. Then, the softmax layer calculates an action probability vector $P = \{P_1, P_2 \cdots P_K\}$, where $K$ denotes the number of types of fitness actions. The action probability vector is calculated as:

$$P = s \left( W^T F + b \right), \quad (5)$$

where $s(\cdot)$ is the softmax function, $W^T$ is a weight matrix and $b$ is a bias vector. $F$ is the output of the previous FC layer. The action label $l$ of the repetition is then assigned to the action with the highest probability

$$l = \arg\max_j P_j, j \in [1, K]. \quad (6)$$

Finally, we add an FC layer and a softmax layer after the third LSTM layer to determine the identity of the user. The softmax layer can calculate an identity probability vector in the same way as the fitness classification module.

We divide the process of training network into two steps: training fitness classification module and training user identification module. First, we use standard fitness data in the fitness template to train the first 2 LSTM layers and the fitness classification module. After training, the first 2 LSTM layers can extract coarse-grained and fine-grained action-level features, and the fitness classification module can distinguish fitness actions in the template. Then, keeping the parameters of the first 2 LSTM layers unchanged, we use the personal fitness data of users to train the last LSTM layer and the user identification module. Note that the fitness classification module can be trained before a user uses our system, while the user identification module can only be trained after collecting certain personal data of users.

To enhance usability, HearFit+ allows users to add new actions besides template. We borrow ideas from incremental learning [19] to make the network recognize new actions. Specifically, the structures and parameters of the LSTM layers are kept unchanged, and the fitness classification module is retrained by the new actions and all existing actions. This method can reduce the size of the data set during retraining. Users only need to complete several repetitions of the new actions. In addition, after a period of exercise, the fitness actions of a user usually have small changes (e.g., more standard). HearFit+ needs to retrain the last LSTM layer and the user identification module to adapt to the changes of fitness actions.

### 3.7 Effect Evaluation

In order to achieve effective fitness, people usually hope that their fitness processes are standard and appropriate. Traditionally, people can invite fitness coaches or partners to observe their fitness processes and provide advice. However, many people who exercise at home/office can not achieve effective fitness. So, the effect evaluation aims to provide users with fine-grained, personalized fitness statistics and intuitive feedback. At present, only a small part of fitness detection research includes effect evaluation [11], [25], [27], [49]. Since these works adopt different devices and algorithms, they define different evaluation metrics. By studying the FITT principle [20] which is a set of guidelines that are commonly used in fitness evaluation, HearFit+ measures fitness qualities from two aspects: local effect and global effect.

#### 3.7.1 Local Effect

The local effect aims to provide users with a more intuitive form to show the completion quality of each repetition. We define 2 metrics named intensity and duration to monitor the quality of each repetition in real-time.

**Intensity.** The intensity reflects the energy expended in each repetition. Each repetition usually consists of two parts: extension and retraction. The balance of their intensities can enhance bidirectional muscle contraction strength and avoid disequilibrium. In HearFit+, we use STE to represent intensity. By observing Fig. 7(b), we can find that the STE of each repetition includes two obvious peaks, representing the intensities of extension and retraction, respectively. We define the balance of intensity as $BI = \alpha - I_e/I_r$, where $I_e$ and $I_r$ are the intensities of extension and retraction, $\alpha$ represents the standard balance obtained from the action template. Finally, we obtain the balance of intensity in each repetition.

**Duration.** The duration reflects the time spent in each repetition. A shorter duration may make the muscles stretch and contract too fierce, which can significantly increase the risk of injury. A longer duration may have a negative effect on muscle flexibility and reaction speed. In addition, the closer to the standard data the duration is, the more standard the repetition is. The duration of each repetition can be directly obtained after the segmentation (Section corresponding standard duration in the action template, we can calculate their difference $d = D_s - D$ as the standard of the duration. The duration is also adopted by other existing solutions [25], [27]. These solutions calculate duration based on peak detection or KL divergence algorithm. Compared with these methods, our slope-based method can calculate the duration more accurately.

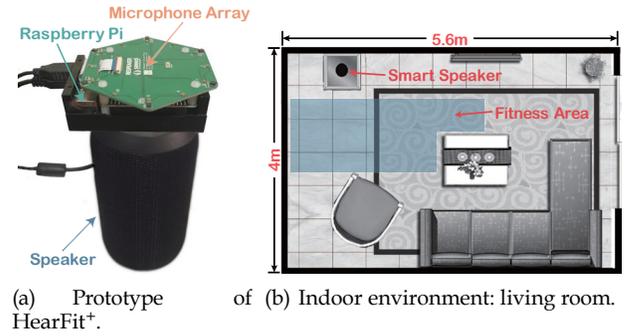

(a) Prototype of HearFit+.  (b) Indoor environment: living room.

Fig. 11. An example of experimental device and environment.

#### 3.7.2 Global Effect

The purpose of the global effect is to show the overall quality of each set of fitness actions. We design two metrics, continuity and smoothness, to reveal the overall stability of the fitness process.

**Continuity.** Continuity reflects the consistency of rest intervals between two repetitions within a set. Effective fitness should maintain a steady rhythm, and unstable rest intervals usually indicate that the intensity of fitness is too high or too low. In order to evaluate the continuity of each set, we adopt the kurtosis as the metric. In probability theory, kurtosis can describe the sharpness of the probability distribution in a real-valued variable. Formally, suppose $R = [r_1, r_2, \cdots, r_n]$ is the rest intervals of a set, the kurtosis can be calculated as:

$$Kurt = \frac{\sum_{i=1}^{n}(r_i - \mu)^4}{\sum_{i=1}^{n}(r_i - \mu)^2} - 3 = \frac{\mu^4}{\theta^4} - 3, \quad (7)$$

where $\mu$ and $\theta$ are the mean and standard deviation of the $R$. A larger $Kurt$ indicates a better continuity of a set.

**Smoothness.** Smoothness reflects the consistency of intensities within a set. The similar intensities of repetitions in a set means that the user can control the muscles well, so as to improve the fitness effect. In order to evaluate the smoothness of each set, the average of $I_e$ and $I_r$ is used as the intensity of each repetition. Similar to continuity, we calculate the intensity kurtosis of a set. And a larger intensity kurtosis indicates a better smoothness of a set.

When users complete the fitness process, they can connect the smartphones to the smart speakers. Then, their fitness statistics in the fitness process are automatically synchronized to their smartphones. According to the intuitive statistics provided by HearFit+, users can improve the fitness effect next time to achieve their fitness goals.

## 4 IMPLEMENTATION AND EVALUATION

In this section, we introduce the implementation details and provide the evaluation results.

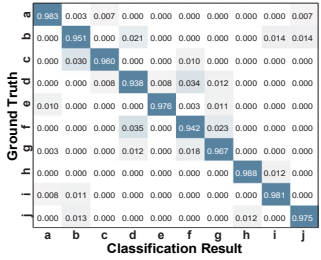

Fig. 12. Overall performance of classification.

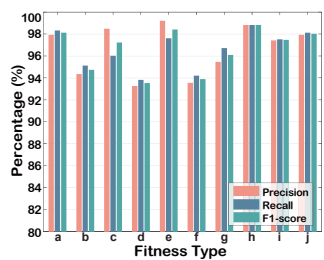

Fig. 13. Precision, recall and F1-score of classification.

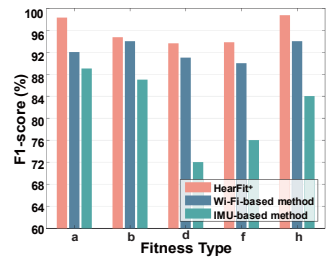

Fig. 14. F1-score of HearFit$^+$ and comparison algorithms.

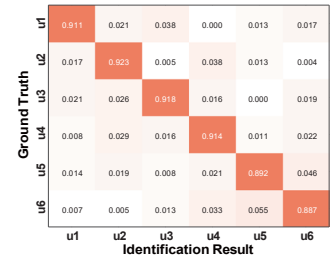

Fig. 15. Overall performance of identification.

TABLE 1
Information of volunteers.

| Attribute \ Gender | Male | Female |
|---|---|---|
| Gender | 9 | 3 |
| Age | 24 - 46 | 17 - 44 |
| Stature | 168 - 186 cm | 161 - 169 cm |
| Weight | 61 - 85 kg | 47 - 56 kg |
| BMI | 20.4 - 25.1 | 18.1 - 19.6 |
| Fitness years | 0 - 8 year | 0 - 4 year |

### 4.1 Experiment Setup

We implement HearFit$^+$ using a Raspberry Pi 4B with a 1.5$GHz$ CPU and 4$GB$ memory, a circular microphone array with 6 microphones, and an omni-directional speaker as shown in Fig. 11(a), which is the same as most commercial smart speakers. The microphone array consists of 6 Knowles omnidirectional microphones and 2 AC108 analog to digital converter (ADC). It is compatible with Raspberry Pi 40-pin headers. The max sample rate it supports is 48$kHz$. The sensitivity of the microphone is $-22dBFS$ and the SNR is 59$dB$. Considering the small size of commercial smart speakers, we choose a distance of 5$cm$ between the microphones to ensure that the prototype is similar in size to commercial smart speakers.

To evaluate the performance of HearFit$^+$, we recruit 12 volunteers, including fitness coaches, people who regularly exercise, and people who rarely exercise, to collect personal fitness data. The information of the 12 volunteers is shown in Table 1. Their fitness process is recorded in 4 different indoor environments, including a bedroom, a living room, a study room, and an office, the size of which are $5.7m \times 3m$, $5.6m \times 4m$, $2.8m \times 3.6m$ and $4.5m \times 4.1m$, respectively. We choose the sizes of the rooms that are most common in residential areas. Fig. 11(b) shows the floor plan of the living room, where the location of the smart speaker is convenient for users to operate. The experiments involve 10 typical types of fitness. During the 4-month experiments, we collect over 9,000 repetitions of all types of fitness for training and evaluation. Specifically, all of the standard fitness data are used to train the fitness classification module. And all of the personal fitness data are used to evaluate the performance of the fitness classification module. We chose half of the personal fitness data of 6 volunteers to train the user identification module. The remaining half of the personal fitness data of the 6 volunteers are used to evaluate the performance of the user identification module. All procedures are approved by the Institutional Review Board (IRB) at Beijing Institute of Technology.

### 4.2 Evaluation Methodology

We mainly evaluate HearFit$^+$ from the following aspects.

**Precision.** The percentage of repetitions that are correctly classified into type/user A in all repetitions classified into A.

**Recall.** The percentage of repetitions that are classified into type/user A in the repetitions truly belong to A.

**F1-score.** It is the harmonic mean of precision and recall. In our system, the F1-score for a specific fitness type/user is defined as $F1 = \frac{Precision * Recall}{Precision + Recall}$.

**Confusion matrix.** Each row and each column of the matrix represent the ground truth and the classification result, respectively. Each entry represents the percentage of a certain fitness type/user that is classified into each class.

### 4.3 Overall Performance

We first evaluate the overall performance of fitness classification based on the bi-functional LSTM network. Fig. 12 shows the confusion matrix of 10 types of fitness actions performed by all volunteers in 4 different environments. HearFit$^+$ achieves an average accuracy of 96.13% for classifying all types of actions. Particularly, we find the dumbbell deadlift and squat have slightly lower accuracy than other types of actions. This is because the two actions mainly contain vertical movements of the upper body, which makes some repetitions be classified to another type. But the lowest accuracy is also over 93.50%.

Fig. 13 shows the corresponding precision, recall, and F1-score of fitness classification. The precision and recall are no less than 93.24% and 93.80%, while the average F1-score is 96.61%. In order to compare the performance of our system with other systems, we select 5 types of fitness actions to compare the F1-score of our system with a Wi-Fi-based method [49] and an IMU-based method [25]. It can be seen clearly from Fig. 14 that HearFit$^+$ performs better than the comparison algorithms. It should be noted that the IMU-based method requires the user to wear additional devices on the upper arm, so this method can not accurately monitor the fitness actions of legs (i.e., dumbbell deadlift, squat, and lunge).

Next, we perform user identification for 4 volunteers who regularly exercise (i.e., u1, u2, u3, and u4) and 2 volunteers who rarely exercise (i.e., u5 and u6). In Fig. 15, the confusion matrix shows the overall accuracy of

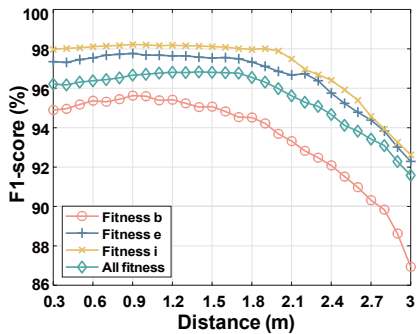
Fig. 16. F1-score of classification under different distances.

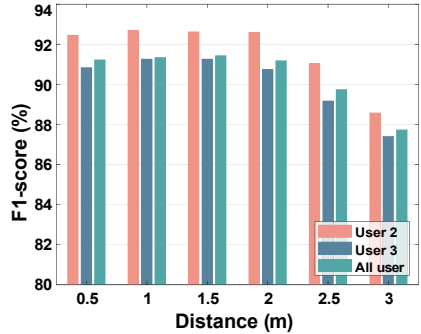
Fig. 17. F1-score of identification under different distances.

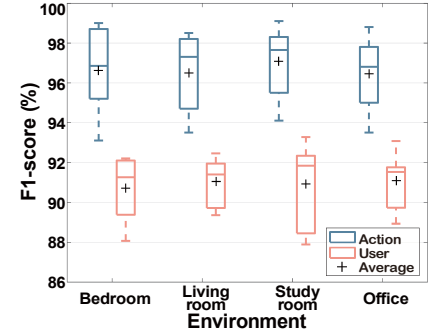
Fig. 18. F1-score under different environments.

TABLE 2
Perceived time delay after user fitness.

| Part  | FD   | NR   | AS   | PFC   | EE  | Total |
|-------|------|------|------|-------|-----|-------|
| Delay | 87ms | 73ms | 17ms | 135ms | 9ms | 321ms |

user identification. We can see that HearFit+ achieves high identification accuracy for all volunteers. Compared with volunteers who regularly exercise, volunteers who rarely exercise lack long-term training, and it is difficult for them to ensure the complete consistency of actions, so our system has a relatively low identification accuracy for them. In addition, after long-term training, fitness actions of users become more and more standard, HearFit+ needs to retrain the last LSTM layer and user identification module to adapt to the changes of fitness actions. It is not clear that the maximum number of users that our system can identify. However, we observe that 6 users are sufficient for most home/office environments.

When the user exercises, our system also processes the received audio signals by Raspberry Pi at the same time. So we define the perceived time delay as the total running time of HearFit+ minus the total fitness time of the user. Table 2 shows the average time delay of *Fitness Detection* (FD), *Noise Reduction* (NR), *Action Segmentation* (AS), *Personalized Fitness Classification* (PFC), and *Effect Evaluation* (EE). It can be seen that the Classification takes the most time, and the total delay time is about $321ms$. In other words, HearFit+ can produce evaluation results within $321ms$ after the user completes the fitness, which indicates that HearFit+ can achieve a satisfactory user experience.

### 4.4 Impact of Different Factors

#### 4.4.1 Impact of Distance

We first study the impact of the distance between the smart speaker and the user. Fig. 16 shows the F1-score of all fitness actions and 3 types of them at different distances. It can be seen that the average F1-score of HearFit+ is more than 94% when the distance is less than $2.4m$, which is sufficient in most indoor environments. In addition, the F1-score of dumbbell curl (Fitness b) decreases relatively faster than others, since only forearms are used in this action and the reflected signal of forearms is relatively weak. Fig. 17 shows the user identification F1-score of all users and 2 specific users with a big difference in body shape. We can find that within $2m$, the distance has little effect on the identification. In addition, the F1-score of the large body (user 2) is slightly higher than that of the small body (user 3). Note that a smaller distance does not always result in a higher F1-score, for HearFit+ can not capture the reflected signal of the user's whole body if the distance is too small. Generally, HearFit+ has good performance when the distance is less than $2m$.

#### 4.4.2 Impact of Environment

Then we evaluate the impacts of four indoor environments that have different furniture densities and layouts, including a bedroom, a living room, a study room, and an office. The distribution of the F1-score for each environment is shown in Fig. 18. It shows that different environments have almost no impacts on fitness classification and user identification as long as there is line-of-sight (LOS) signal. Moreover, when there are small objects between users and smart speakers, such as benches, boxes, or dustbins, the classification F1-score of sit-up slightly decreases by 0.4% and the identification F1-score decreases by 0.9%. This may be because when the user lies on the floor, the small objects block some reflected signal of the user's body, while only block little signal when he/she takes other actions.

#### 4.4.3 Impact of Interference

The reflected signal of the user would be affected by surrounding activities. We study the impacts on fitness classification when there is someone walking, typing, watching TV, listening to music, and a pet wandering around the user, in terms with and without noise reduction (Section 3.4). From Fig. 19, we can see that F1-score is relatively low when there is a walking person, and the closer to the user the walking person is, the lower the F1-score is. But according to proxemics [50], the social distance between people is larger than $2m$. Thus, people barely get too close to the user. Moving pets and people who are typing on a keyboard also have slight impacts. Since these movements are tiny or closer to the floor, we can use noise reduction to filter them out. Watching TV and listening to music have almost no impacts on HearFit+. Since the sounds of TV and music mainly have low-frequency components while HearFit+ emits and collects high-frequency signal. For similar reasons, these interference factors have limited impacts on user identification. No matter in which condition, it is obvious that noise reduction significantly improves the performance of HearFit+.

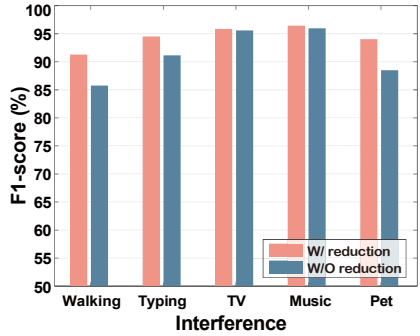

Fig. 19. F1-score of classification under different interferences.

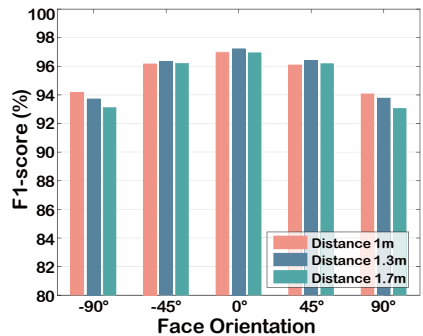

Fig. 20. F1-score of classification under different orientations.

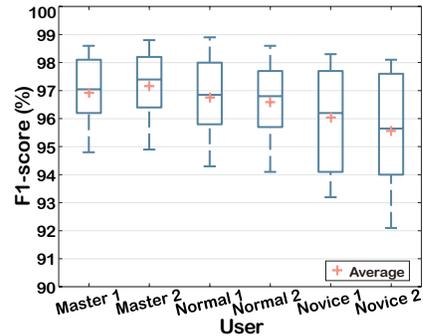

Fig. 21. F1-score of classification under different fitness levels.

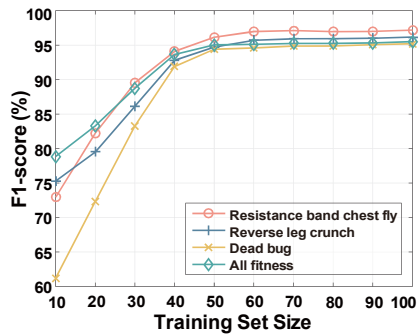

Fig. 22. F1-score for new actions.

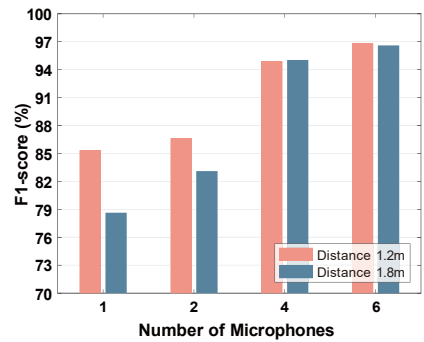

Fig. 23. F1-score under different microphone number.

#### 4.4.4 Impact of Face Orientation

It is possible for users to face different orientations in different environments. We define the degree as $0°$ when a user completely faces the smart speaker. When the user turns left, the degree decreases, otherwise the degree increases. We measure the classification F1-score of 5 angles at 3 distances, and the results are shown in Fig. 20. The F1-score of all angles exceeds 93%, and it keeps relatively high when the angle is between $-45°$ and $45°$. If the angle exceeds this range, the F1-score gradually decreases. A special condition is that when the user turns back to the smart speaker, the movements of some parts of his/her body (e.g., forearm) can not be detected by HearFit+. So, in order to achieve better performance of classification and identification, we suggest that users face their smart speakers as much as possible.

#### 4.4.5 Impact of Fitness Level

6 volunteers are involved in this experiment with height ranging from $1.61m$ to $1.86m$ and weight ranging from $47kg$ to $82kg$, including fitness coaches, people who regularly exercise, and people who rarely exercise. We divide their fitness levels into 3 classes: master, normal, and novice. Fig. 21 shows the distribution of classification F1-score for 6 volunteers at different levels. We can see that HearFit+ has good performance on both master and normal users. However, due to the fact that novices have almost no fitness training, they can not keep stable when doing some actions (e.g., one leg deadlift and lunge), leading to a slightly lower F1-score. But after a few training, the F1-score of novice users can reach the normal level. We observe similar processes in user identification for novice users, HearFit+ needs to retrain the last LSTM layer and the user identification module regularly to adapt to the changes of users' actions. On a whole, HearFit+ performs well under different fitness levels and different body types.

#### 4.4.6 Impact of New Action

When users add new actions, HearFit+ needs to retrain the fitness classification module. Thus, we study how many repetitions of new and previous actions should be used to achieve a good performance. We add 3 new actions that are reverse leg crunch, dead bug, and resistance band chest fly. Fig. 22 shows the F1-score under different training set sizes of each action. We can see that when the training set is greater than 50, the average F1-score of all fitness actions is greater than 95%. In other words, to add a new action, HearFit+ only needs to collect 50 repetitions. In addition, when the user identification module is retrained, the data of new action are also added to the training set. Compared with retraining all layers, this method based on incremental learning can reduce training data by 82% and training time by 96%.

#### 4.4.7 Impact of Microphone Number

The number of microphones affects the performance of beamforming. Here, we quantitatively analyze the impact of microphone number on fitness classification at 2 distances.

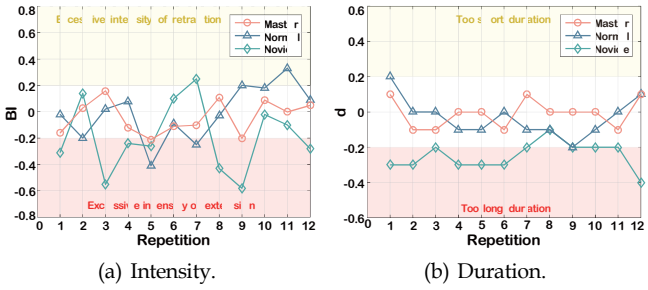

(a) Intensity.   (b) Duration.

Fig. 24. Local effect of 3 volunteers.

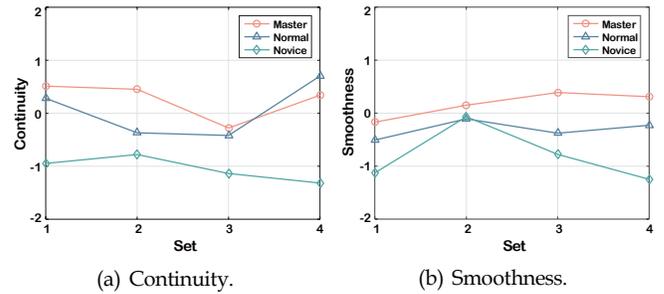

(a) Continuity.   (b) Smoothness.

Fig. 25. Global effect of 3 volunteers.

The volunteers completely face the smart speaker and perform fitness actions. We change the distances between the smart speaker and the volunteers and collect the data on the smart speaker. We then extract the reflected signal using a) only a single microphone on the smart speaker without beamforming; b) 2 microphones on the left and right of the smart speaker; c) 4 microphones in corners of the smart speaker; and d) all 6 microphones to process the signal.

The F1-score of classification results is shown in Fig. 23. It shows that multiple microphones can improve the detection range and accuracy. HearFit+ with a single or 2 microphones can reach an F1-score of 85% at a distance of $1.2m$, which means that HearFit+ is not only deployed on smart speakers, but also has great potential to be deployed on smartphones. When HearFit+ uses 4 or more microphones, and the range is extended to $1.8m$, the F1-score is improved to 95%. At present, most commercial smart speakers have more than 3 microphones, so HearFit+ is suitable for deployment on most smart speakers.

### 4.5 Performance of Effect Evaluation

In order to test the performance of effect evaluation, we ask 3 volunteers at different levels to do 4 sets of squats with each set containing 12 repetitions. We measure the quality from aspects of local effect and global effect.

#### 4.5.1 Local effect

Fig. 24(a) shows the intensity of each repetition in the last set. It can be seen that the master user can maintain a relatively stable and standard intensity. The normal user maintains an appropriate intensity in the first half of the set. Due to a lack of physical strength in the last set, the normal user begins to lose stability in the last few repetitions. The novice user has the most unstable intensity due to lacking fitness training. But after about two months of training, his/her actions can become standard. Fig. 24(b) shows the duration of each repetition in the last set. Similar to intensity, most repetitions of the master user and the normal user have stable duration, and most of the repetitions in the set can maintain a standard duration. The novice user can also maintain a relatively stable duration, but most of the durations of the novice user are larger than the standard duration.

#### 4.5.2 Global effect

Finally, we evaluate the global effect of the 4 sets. The closer to 0 the value of intensity and duration is, the better the local effect is, while the larger the values of continuity and smoothness are, the better the global effect is. Fig. 25(a) shows the continuity of each set. It can be seen from the figure that the master user has the best continuity, and the continuity of the normal user is only slightly lower than that of the master user. When the number of sets increases, the continuity of the novice user decreases significantly. Since novice user does not participate in fitness training, their physical strength decrease during the third and fourth sets, and they can not maintain stable fitness action. The same trend appears in smoothness evaluation, as shown in Fig. 25(b). For the novice user, the smoothness is better than the continuity, which means that the user tries to maintain the stability of each repetition, but can not guarantee the stability of rest intervals. When users view fitness statistics, we can provide them with standard fitness data as a reference. They can improve the fitness effect according to the statistics provided by HearFit+.

TABLE 3
The mean and variance of the 4 metrics for 3 users.

| Level | Intensity | | Duration | | Continuity | | Smoothness | |
|---|---|---|---|---|---|---|---|---|
| | Mean | Var | Mean | Var | Mean | Var | Mean | Var |
| Mas | 0.018 | 0.016 | 0.008 | 0.007 | 0.255 | 0.132 | 0.17 | 0.061 |
| Nor | 0.013 | 0.043 | 0.033 | 0.010 | 0.047 | 0.291 | -0.307 | 0.03 |
| Nov | 0.19 | 0.072 | 0.25 | 0.009 | -1.047 | 0.104 | -0.807 | 0.281 |

In order to objectively evaluate the fitness quality of users, we calculate the mean and variance of the 4 metrics. Table 3 shows the mean and variance of the 4 metrics for 3 users. The closer to 0 the mean of intensity and duration is, the better the local effect is, while the larger the mean of continuity and smoothness is, the better the global effect is. We can see that the usual order of fitness effects is master > normal user > novice. The variance close to 0 indicates good fitness stability. Considering the 4 metrics, the master has the best stability, while the novice has the worst stability. We also ask fitness coaches and master users to perform several sets of fitness actions. Finally, they view the fitness statistics and compare the statistics with their actual feelings. Overall, all volunteers give positive feedback on effect evaluation and confirm that HearFit+ can help users to improve fitness effect.

## 5 DISCUSSION AND LIMITATIONS

We evaluate several factors that affect HearFit+. But there are also several limitations and opportunities to improve it.

**NLOS conditions.** According to the acoustic theory, the higher the frequency of sound is, the lower the penetration rate is. And in the same medium, the higher the sound frequency is, the faster the sound attenuates. So, compared with the low-frequency signal, the ability of the ultrasonic signal to penetrate walls is relatively poor. In addition, the ultrasound can bypass the obstacle if the size of the obstacle is similar to the wavelength of the ultrasound. The average wavelength of the signal we used is about $1.7cm$, which is much smaller than the sizes of most of the obstacles usually presented in a room. So HearFit+ is suitable to work under LOS conditions with no obstacles between most of the user's body and the smart speaker. As our future work, we will use multiple smart speakers at different locations to overcome NLOS signals.

**Multiple users.** Our system can monitor the fitness activities of one user at a time. But when multiple users exercise at the same time, our system can hardly handle this scenario well. At present, multi-user action detection is mainly used for breathing and heartbeat, which can remain stable for a long time. However, during fitness, users often change their fitness actions, and multiple users are likely to perform the same fitness actions, which also brings great challenges to multi-user fitness monitoring. As our future work, we will further research related technologies (e.g., ZC sequences) to extend our system for multi-user cases.

**Non-stop fitness actions.** We design the segmentation algorithm of HearFit+ based on the fact that people tend to take a short rest between consecutive repetitions to control the training pace for most fitness actions. But there are still some fitness actions without obvious rest, such as jumping jack and bobby jump. When using the segmentation algorithm to deal with these non-stop actions, the accuracy drops to a certain extent. In our future work, we will design a more universal segmentation method for these kinds of fitness actions.

# 6 CONCLUSION

In this paper, we design and implement HearFit+, the first non-invasive personalized fitness monitoring system based on smart speakers working at home/office, which aims to provide users with fine-grained fitness statistics to improve the fitness effect. HearFit+ extracts features from fitness actions based on Doppler profiles, then adopts a bi-functional LSTM network to classify fitness actions and identify users. Finally, it evaluates fitness effect from the local view and the global view. HearFit+ is anti-interference by adopting beamforming, it also allows users to add new actions conveniently. The extensive experiments show that it achieves an average classification accuracy of 96.13% and identification accuracy of 91%, and can provide accurate fitness statistics.

## ACKNOWLEDGMENTS

The work of Fan Li is partially supported by the Beijing Natural Science Foundation under Grant No. 4192051, and the National Natural Science Foundation of China (NSFC) under Grant No. 62072040, 61772077.

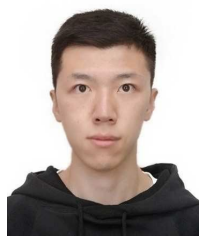

**Yadong Xie** received the BE degree in network engineering from Hebei University, China in 2016. Currently he is a Ph.D. candidate in the School of Computer Science, Beijing Institute of Technology, Beijing, China. His research interests include mobile computing, mobile health, human-computer interaction, and deep learning.

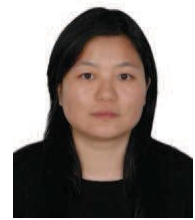

**Fan Li** received the PhD degree in computer science from the University of North Carolina at Charlotte in 2008, MEng degree in electrical engineering from the University of Delaware in 2004, MEng and BEng degrees in communications and information system from Huazhong University of Science and Technology, China in 2001 and 1998, respectively. She is currently a professor at School of Computer Science in Beijing Institute of Technology, China. Her current research focuses on wireless networks, ad hoc and sensor networks, and mobile computing. Her papers won Best Paper Awards from IEEE MASS (2013), IEEE IPCCC (2013), ACM MobiHoc (2014), and Tsinghua Science and Technology (2015). She is a member of ACM and IEEE.

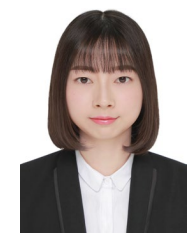

**Yue Wu** received the BE degree in Internet of things from Beijing Institute of Technology, China in 2015. Currently she is a Ph.D. candidate in the School of Computer Science, Beijing Institute of Technology, Beijing, China. Her research interests include mobile crowd sensing, edge computing, and deep learning.

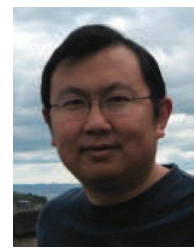

**Yu Wang** is currently a Professor in the Department of Computer and Information Sciences at Temple University. He holds a Ph.D. from Illinois Institute of Technology, an MEng and a BEng from Tsinghua University, all in Computer Science. His research interest includes wireless networks, smart sensing, and mobile computing. He has published over 200 papers in peer reviewed journals and conferences, with four best paper awards. He has served as general chair, program chair, program committee member, etc. for many international conferences (such as IEEE IPCCC, ACM MobiHoc, IEEE INFOCOM, IEEE GLOBECOM, IEEE ICC). He has served as Editorial Board Member of several international journals, including IEEE Transactions on Parallel and Distributed Systems. He is a recipient of Ralph E. Powe Junior Faculty Enhancement Awards from Oak Ridge Associated Universities (2006), Outstanding Faculty Research Award from College of Computing and Informatics at the University of North Carolina at Charlotte (2008), Fellow of IEEE (2018), and ACM Distinguished Member (2020).